\documentclass[pra,showpacs,twocolumn]{revtex4}
\usepackage{amsmath}
\usepackage{amsfonts,amssymb,graphicx}

\begin{document}

\title{ Covariant feedback Control of arbitrary qudit state against Depolarizing Noise is impossible }
\author{ShengLi Zhang$^{1,2}$, XuBo Zou$^{1}$, Chuanfeng Li$^{1}$,ChenHui
Jin$^{2}$, and GuangCan Guo$^{1}$} \affiliation{1 Key Laboratory of
Quantum Information, University of Science and
Technology of China (CAS), Hefei 230026, China.\\
2 Electronic Technology Institute; Information Engineering University;
Zhengzhou; Henan 450004; China}
\date{\today}

\begin{abstract}
In this paper, we prove that it is impossible to construct a
convariant quantum feedback control instruments which aim to correct
the channel noise imposed on an unknown qubit. The proof is based on
the searching for the optimal quantum control protocol and it
turns out there exist no better %\textit{Complete Positive Trace
%Preserving }(CPTP) maps
complete positive and covariant quantum operations which provide a
higher fidelity than the trivial Identity operators. The
generalization of the investigation to bipartite entangled pure
state is also included.

\end{abstract}

\maketitle

The laws of quantum mechanics impose a number of \textit{no-go}
theorems on many Quantum Information Processing. Examples of such
impossible tasks can be provided by the famous no-cloning
theorem\cite{qcloNature,qcloPLA}, no-deletion
principle\cite{qdelNature} or by the theorem on nonexistence of the
universal-NOT gates\cite{cnot99}. Despite its discouraging
impossibility, lots of attention have been paid to construction of
the optimal and approximate quantum instruments for
duplicating\cite{qcloRMP}, deleting\cite{qdelPRA} and
complementing\cite{cnot00} an unknown quantum state. These
considerations are of great interest since they provide great
insight into the fundamental limits on the distribution and
manipulation of quantum information and can become of practical
relevance for analyzing the security of quantum cryptography and
quantum communications.

In this work, we give an investigation of quantum feedback control
of a completely unknown qubit and provide another \textit{no-go}
theorem. Our main result is that, by driving the optimal and
covariant correction maps for correcting errors of the completely
unknown quantum pure state which was previously subjected to the
depolarizing noise, there is no better other quantum operations maps
that will provide a higher fidelity than the trivial Identity
operators.

In the literature, counteracting the effects on quantum systems of
noise imposed by environment is also a central problem for quantum
information technology. In 1983, V. P. Belavkin investigated the
feedback control of quantum system and proposed a theoretic
framework for control model\cite{Bel83}. Since this seminal work,
the model of quantum feedback control has received a wide and
extensive attention\cite{Bel04,ctlQbit,JWang,Pryde,
Wiseman,OpeErrCor,Barnum02,Tico,Lidar,R.van,Greg04}. The methods
utilized in these models can be generally divided into two
categories. One is to investigate the continuous time feedback,
which includes: the stabilization of a single state of a driven and
damped two-level atom\cite{JWang, Wiseman}, the maintenance of the
coherence of a noisy qubit using tracking control\cite{Lidar}, the
state preparation and stabilization onto eigenstates of a continous
measuremed observable in higher-dimensional system\cite{R.van}, and
more recently, the correctability of quantum subsystems under
continuous dynamics evolution has also been studied\cite{OpeErrCor}.
The other is to study the problem in a discrete time setting which
considerably simplifies the problem and most clearly illustrates the
essential concepts. For example, in Ref.\cite{Barnum02}, Barnum and
Knill gave a near-optimal correction procedure to recover the
ensembles of orthogonal states from a general noisy process. Similar
problems have also been considered in Ref.\cite{Tico} where Ticozzi
\textit{et.al} considered the suppression of the unwanted dynamics
of a single qubit by applying both dynamic decoupling and feedback
methods and in Ref.\cite{Greg04} where the criteria for quantum
information to be perfectly corrigible and the related feedback
operation is obtained. Moreover, experiments set-up on quantum
control of a single photonic qubit, have already been
reported\cite{Pryde}.

The quantum feedback control problem is typically the following.
Suppose a pure quantum state is initially prepared and later, damped
and disturbed by the environmental noise. One performs some kinds of
quantum operations such as measurement and feedback control on the
damped state to achieve optimal or near-optimal preservation of the
initial pure state. Particularly, in Ref. \cite{ctlQbit}, the
optimal control of a qubit against the dephasing channel noise is
considered. It was shown there that one can apply a measure and
feedback-control quantum operation and further correct, at least to
some extent, the unknown qubit if it was originally prepared in one
of two non-orthogonal states and subsequently subjected to the
dephasing noise. However, in this paper, we find that the
construction of a similar depolarizing-noise-counteracting feedback
control is impossible, for a completely unknown quantum pure state
which resides in the arbitrary $D$-dimensional Hilbert Space
$\mathcal{H}_D$.

The structure of our paper is organized as follows. We will first
give a brief introduction to the quantum depolarizing channel and
also the figure of metric which characterizes the optimality of our
noise-counteracting maps. Then, we will exploit the group symmetry
property and obtain the optimal covariant control of the unknown
qudit. Finally, following such a framework, we also give a
generalization from the single party quantum state to the bipartite
 pure state, and consider the optimal quantum recovery of maximal entanglement.

\begin{figure}
  % Requires \usepackage{graphicx}
  \includegraphics[width=7cm]{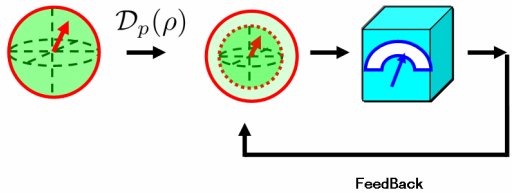}\\
  \caption{Feedback control of single qudit. Suppose a pure quantum state is initially prepared and later, damped
and disturbed by the environmental noise. One performs some kinds of
feedback control operations to counteract the channel noise.
}\label{fig1}
\end{figure}

The noise model we are interested in is the depolarizing channel
which is an important type of quantum noise\cite{NielsenChuang}.
Imagine we randomly take an arbitrary state $\rho$, from the entire
Hilbert Space $\mathcal{H}_D$, and with a probability $p$ that the
qubit is depolarized. Described by the \textit{Complete Positive
Trace Preserving }(CPTP) map $\mathcal{D}_p$, the noise can be
characterized by\cite{NielsenChuang}:
\begin{equation} \mathcal{D}_p(\rho)=p\frac{I}{D}+(1-p)\rho.
\end{equation}
Our main task at hand is to search for a quantum correction
procedure $\cal{C}$ which measures the disturbed state and later
feedback-controls the state to preserve the initial pure state
optimally. We give a description of the measurement and controlling
scheme in Fig.\ref{fig1}.

To continue our discussion, we need a figure of metric that
quantifies the performance of our control
operation. For ease, we will use the Uhlmann's fidelity\cite{Uhl76}.
This is a good evaluation of the resemblance of two quantum state.
For the original pure state $\rho=|\psi\rangle\langle\psi|$ and the
corrected state $\rho'=\mathcal{C}(\mathcal{D}_{p}(\rho))$, the
fidelity can be given by:
$F_\rho=\langle\psi|\mathcal{C}(\mathcal{D}_{p}(\rho))|\psi\rangle$.
Thus, the performance of the correction map $\mathcal{C}$ can be
formulated by averaging over all the possible input pure state:
\begin{eqnarray}\label{fave}
F=\int_\rho d_\rho F_\rho=\int_\rho
d_\rho\langle\psi|\mathcal{C}(\mathcal{D}_{p}(\rho))|\psi\rangle.
\end{eqnarray}

In the following discussions, we assume that we have no a priori
knowledge about the unknown qudit. Such a state in two-level system
can be represented with the orthogonal basis
$\{|0\rangle,|1\rangle\,\dots,|D-1\rangle\}$ and automatically forms
an orbit of the group $\mathbb{G}=SU(D)$ of $D\times D$ matrices
with the determinant $+1$. Thus, each state on the Bloch sphere can
be represented with an element of the group $\mathbb{G}$:
\begin{eqnarray}
|\psi_g\rangle=U_g|0\rangle,
\end{eqnarray} in which $U_g$ is the Unitary
Representation of $SU(D)$.

In the view of no \textit{a priori} knowledge, one can expect the
initial probability distribution is uniform with respect to
\textit{Haar }measure $dg$ in the ensemble average
(Eq.(\ref{fave}))\cite{Cornwell}. So the average fidelity follows
\begin{eqnarray}
F=\int_{SU(D)}dg\langle\psi_g|\mathcal{C}[\mathcal{D}_p(|\psi_g\rangle\langle\psi_g|)]|\psi_g\rangle.\label{fid}
\end{eqnarray}

The problem is to look for the optimal operation $\cal{C}$ which
exhibits the best possible performance. According the Kraus's
representation\cite{Kraus}, an arbitrary quantum operation
$\mathcal{C}$ can be represented with a set of CPTP maps
$\{\mathcal{C}_h\}$ with $h$ indicating the operation outcome.
Furthermore, written with Kraus operator, each map can then be
written with $\mathcal{C}_h(\rho)=\sum_{\mu}A_{h\mu}\rho
A_{h\mu}^{\dagger}$ and can provide the result $h$ with a
probability $p_h=\mathrm{Tr}\left[\rho
\sum_{\mu}A_{h\mu}^{\dagger}A_{h\mu}\right]$. Note that the overall
map $\mathcal{C}$ must be trace preserving, which imposes the
constraints that
\begin{eqnarray}\label{tra1759}
 \int_h dh \sum_{\mu}A_{h\mu}^{\dagger}A_{h\mu}=I.
\end{eqnarray}

In general, optimization of the quantum operation $\cal{C}$ is
equivalent to search for the optimal operation $\{A_{h\mu}\}$ over
all CPTP maps acting on a single qubit. This is quite intractable in
practice. However, by introducing an isomorphism and utilizing the
group covariance property of our problem, we will find such a
problem can be evaluated as a maximal eigenvalue of certain
operator. In Ref.\cite{Jam}, Jamio\l kowski established an
isomorphism between the completely positive map $\mathcal{C}_h$ on
Hilbert Space $\mathcal{H}$ and positive semi-definite operators
$R_h$ on $\mathcal{H}_{out}\otimes\mathcal{H}_{in}$. Such a
correspondence is one-to-one and its inverse can be given by
\begin{eqnarray}
\mathcal{C}_h(\rho)=\mathrm{Tr}_{in}[(I_{out}\otimes\rho^{\tau})R_h],
R_h=(\mathcal{C}_h\otimes
I)|\mathcal{I}\rangle\langle\mathcal{I}|,\label{map}
\end{eqnarray}
in which the relevant Hilbert Space $\mathcal{H}$ is $D$-dimensional
and $|\mathcal{I}\rangle=\sum_{i}|i\rangle_{out}|i\rangle_{in}$ is
the unnormalized maximal entangled state in
$\mathcal{H}_{out}\otimes\mathcal{H}_{in}$.

With such an isomorphism, the trace preserving constraints for $R_h$
should be
\begin{eqnarray}
\mathrm{Tr}_{out}[\int_h dh R_h]=I_{in}, \label{trace}
\end{eqnarray}
and the average fidelity in Eq.(\ref{fid}) follows:
\begin{eqnarray}
F=\int dg\int dh
\mathrm{Tr}[(\rho_g\otimes\mathcal{D}_p(\rho_g)^\tau)R_h],\label{fid2}
\end{eqnarray}
where $\rho_g=|\psi_g\rangle\langle\psi_g|$.

To move on, we will consider the covariant quantum measurement
\cite{Helstrom76,Davies,Extremal}. The term \textit{covariant}
implies that the performance of the measurement does not depend on
the specific choice of input state and has been proven to be optimal
in state estimation\cite{qes} and quantum cloing\cite{qclo} process.
We observe that such an optimality is also true with respect to our
quantum control problem. This observation allows us to greatly
reduce the complexity of the optimization. In fact, for an arbitrary
non-covariant quantum measurement $\{R_h\}$ with the average
fidelity $F$, one can also construct a covariant instrument
$\{\tilde{R}_h\}$:
\begin{eqnarray}
\tilde{R}_0&=&\int_{\mathbb{G}} dh (U_h^{\dagger}\otimes
U_h^{\tau})R_h
(U_h\otimes U_h^{*}),\\
\tilde{R}_{h}&=&(U_h\otimes U_h^{*})\tilde{R}_0(U_h^{\dagger}\otimes
U_h^{\tau}),\label{COV}
\end{eqnarray}
which preserve the same average fidelity as $\{R_h\}$, i.e.,
$\tilde{F}=F$. Thus, the optimal average fidelity for covariant
quantum control $\tilde{R}$ is also the optimal bound for
non-covariant maps. Therefore, in looking for the optimal quantum
control, there will be no loss of generality if we restrict our
search within covariant maps.

In the rest of the paper, we will omit the tilde $"\sim "$ for
clearance. By considering the covariant map in Eq.(\ref{COV}), the
fidelity in Eq.(\ref{fid2}) can be given by
\begin{eqnarray}
F&=&\int_{\mathbb{G}} dg
\mathrm{Tr}[\rho_g\otimes\mathcal{D}_p(\rho_g)^\tau R_0]\nonumber\\
&=&\mathrm{Tr}[\Upsilon R_0],\label{fid3}
\end{eqnarray}
where $\Upsilon=\int_{\mathbb{G}} dg
\rho_g\otimes\mathcal{D}_p(\rho_g)^{\tau}$ and can be easily
calculated. Actually,
\begin{eqnarray}
\Upsilon &=& \int_{\mathbb{G}} dg
\rho_g\otimes\mathcal{D}_p(\rho_g)^{\tau}\nonumber\\
&=&\int_{\mathbb{G}} dg U_g\otimes
U_g^{*}\left[|0\rangle\langle0|\otimes\left(p\frac{I}{D}+(1-p)|0\rangle\langle0|\right)\right]
U_g^{\dagger}\otimes U_g^{\tau}.\nonumber
\end{eqnarray}
Using Schur's lemma for reducible group representations \cite{zelo},
it can be obtained that
\begin{eqnarray}
\Upsilon =
\frac{1-p}{D(D+1)}|\mathcal{I}\rangle\langle\mathcal{I}|+\frac{D+p}{D^{2}(D+1)}I\otimes
I,
\end{eqnarray}
where $|\mathcal{I}\rangle=\sum_{i=0}^{D-1}|i\rangle|i\rangle$.

Similarly, following the Schur's lemma for irreducible group
representations, we have the the identity $\int_{\mathbb{G}} dg
U_{g}X U_{g}^{\dagger} =\frac{1}{D}\mathrm{Tr}[X]I$ and the
condition Eq.(\ref{trace}) can be equivalently reduced to
\begin{eqnarray}\label{trace2}
\mathrm{Tr}[R_0]=D.
\end{eqnarray}

The optimal fidelity can now be easily cast to looking for a
positive operator $R_0$ which satisfies Eq.(\ref{trace2}) and
maximizes $F$. Then, $R_0$ will provide a covariant instrument that
achieves the optimal feedback control. It follows that the fidelity
is bounded above by the maximum eigenvalue $\lambda_{max}$ of the
operator $\Upsilon$. One can check that, for any value of $p$, the
maximum eigenvalue and the corresponding eigenvector is given by
\begin{eqnarray}
\lambda_{max}=\frac{D(1-p)+p}{D^2},
|\lambda_{max}\rangle=\frac{1}{\sqrt{D}}|\mathcal{I}\rangle,
\end{eqnarray}
Thus, the optimal operator $R_0$ can be chosen as
$R_0^{(opt)}=|\mathcal{I}\rangle\langle\mathcal{I}|$ and the
corresponding fidelity is $F^{(opt)}=1-p+\frac{p}{D}$.

From Eq.(\ref{map}), it follows that the optimal feedback control
can be achieved by an instrument with Kraus operators $A_g=I$, which
denotes the trivial identity maps and confirms us that the optimal
reversal and feed-back control is to do nothing. That is to say,
there exist no better complete positive and covariant quantum
operations which provide a higher average fidelity.

Before concluding our paper, it should also be noted that such a
framework of deriving optimal covariant feedback control of single
partite pure state can be directly generalized to the bipartite
entangled cases. Bipartite entanglement state, especially, the
maximal entanglement state, is a fundamental resource for quantum
teleportation, quantum cryptography and for quantum
communication\cite{frame}. Consider the following problem: Alice,
who owns a pair of maximally entangled state $|\Psi\rangle_{AB}$,
wants to establish some quantum entanglement by sending one of the
entanglement particles $B$ to remote Bob via the noisy quantum
channel. A question that naturally arises in this field is whether
there exist nontrivial CP maps that could increase the amount of
shared entanglement.

In this following, let's simulate such a problem with a maximal
entanglement state which is randomly chosen from  $\mathcal{H}_D
\otimes\mathcal{H}_D$ and with the depolarizing channel. Using the
notation in Ref. \cite{GMD}, the set of all the possible maximally
entangled state can also be generated from the fixed state
$\frac{1}{\sqrt{D}}|\mathcal{I}\rangle$:
\begin{eqnarray}
\Omega=\{|\Psi_g\rangle_{AB}=\frac{1}{\sqrt{D}}U_g\otimes
I|\mathcal{I}\rangle,U_g \in SU(D)\}.
\end{eqnarray}
The depolarizing channel, which takes effect only on the second
particle $B$, can be formulated by
\begin{eqnarray}
\mathcal{D}_p^{(B)}(|\Psi_g\rangle\langle\Psi_g|)=U_g\otimes I
\left(\frac{1-p}{D}|\mathcal{I}\rangle\langle\mathcal{I}|+\frac{p}{D^2}I\right)U_g^{\dagger}\otimes
I.\nonumber\\~
\end{eqnarray}
Without loss of generality, we still consider the covariant operator
$\{R_h^{(ent)}\}\in
\mathcal{H}_{out}\otimes\mathcal{H}_{in}\triangleq
\mathcal{H}^{(1)}\otimes\mathcal{H}^{(2)}\otimes\mathcal{H}^{(3)}\otimes
\mathcal{H}^{(4)}$:
\begin{eqnarray}
R_{h}^{(ent)}=(U_h^{(1)}\otimes
U_h^{*(3)})R_0^{(ent)}(U_h^{\dagger(1)}\otimes U_h^{\tau(3)})
,\label{COVent}
\end{eqnarray}
with the trace preserving condition $\mathrm{Tr}[R_0]=D^2$. In this
way, we have the average fidelity:
\begin{eqnarray}
F^{(ent)}&=&\int_{\mathbb{G}} dh
\mathrm{Tr}[(|\Psi_h\rangle\langle\Psi_h|\otimes\mathcal{D}_p^{(B)}(|\Psi_h\rangle\langle\Psi_h|)^\tau)R_0^{(ent)}]\nonumber\\
&=&\mathrm{Tr}[\Xi R_0^{(ent)}],\label{fidEnt}
\end{eqnarray}
with $\Xi$ is a integral which can be still straighfordly evaluated
following the reducible group representation\cite{zelo}:
\begin{eqnarray}
\Xi &=&
\frac{D^{2}-p}{D^{4}(D^{2}-1)}I+\frac{1-p}{D^{2}(D^{2}-1)}|\mathcal{I}\rangle_{13}\langle\mathcal{I}|\otimes
|\mathcal{I}\rangle_{24}\langle\mathcal{I}|\nonumber\\
&-&\frac{1-p}{D^{3}(D^{2}-1)}\left(I^{(13)}\otimes|\mathcal{I}\rangle_{24}\langle\mathcal{I}|
+|\mathcal{I}\rangle_{13}\langle\mathcal{I}|\otimes
I^{(24)}\right).\nonumber\\~
\end{eqnarray}

The operator $\Xi$, whose maximal eigenvalue determines the maximal
achievable fidelity can be easily calculated and we have
$\lambda_{max}^{(ent)}=\frac{1-p}{D^2}+\frac{p}{D^4}$ and
$|\lambda_{max}^{(ent)}\rangle=\frac{1}{D}|\mathcal{I}\rangle_{13}|\mathcal{I}\rangle_{24}$.
However, the maximal fidelity $F^{(ent)}_{opt}=1-p-\frac{p}{D^2}$ is
no larger than the case when no quantum recover operation is done.
This coincides with the fact that the optimal quantum operation is
$R_{0,opt}^{(ent)}=D^{2}|\lambda_{max}^{(ent)}\rangle\langle\lambda_{max}^{(ent)}|$
and in Kraus's form we have $A_g=I^{(AB)}$.

In summary, by means of group covariant technology, we give a
thorough and analytically proof of the existence of the nontrivial
control schemes for depolarizing channel. The quantum states we
investigated include both single quantum state and bipartite quantum
entangled state. However, it should be noted that whether these
results can be generalized to the multi-copy quantum state is still
open. This is quite an interesting but relative intricate problem,
which may require a further and systematic investigation in the
future.

 This work was
supported by National Fundamental Research Program, also by National
Natural Science Foundation of China (Grant No. 10674128 and
60121503) and the Innovation Funds and \textquotedblleft Hundreds of
Talents\textquotedblright\ program of Chinese Academy of Sciences
and Doctor Foundation of Education Ministry of China (Grant No.
20060358043).

\end{document}